\documentclass[a4paper,12pt]{article}
\font\Sets=msbm10

\def\Real{\hbox{\Sets R}}

\def\bb{\begin{equation}}
\def\ee{\end{equation}}

\def\mod{\hbox{mod}}
\def\pt{\partial}
\def\const{\hbox{const}}
\def\th{\hbox{tanh}}
\def\ch{\hbox{cosh}}

\def\a{\alpha}

\def\G{\Gamma}
\def\g{\gamma}

\def\l{\lambda}
\def\t{\tau}
\def\ve{\varepsilon}
\def\s{\sigma}

\title{Asymptotics of soliton solution for the perturbed
Davey-Stewartson-1 equations\footnote{This work was
supported by RFBR 97-01-00459}}
\author{O.M.Kiselev,  \\
Institute of Mathematics,\\ Ufa Sci Centre of Russian Acad. of Sci\\
112, Chernyshevsky str., Ufa, 450000, Russia\\ E-mail: ok@
 imat.rb.ru}
\date{October 14,  1999}

\begin{document}
\maketitle
\begin{abstract}
The dromion of the Davey-Stewartson-1 equation is studied
under perturbation on  the large time.
\end{abstract}

In this work we construct the asymp\-to\-tic
so\-lu\-ti\-on of the Da\-vey-Ste\-wart\-son-1 equations
(DS-1):
\begin{eqnarray}
i\pt_t Q+{1\over2}(\pt_\xi^2+\pt_\eta^2)Q+(G_1+G_2)Q=\ve
iF,
\nonumber\\
\pt_\xi G_1=-{\s\over2}\pt_\eta |Q|^2,\quad
\pt_\eta G_2=-{\s\over2}\pt_\xi |Q|^2.
\label{ds1}
\end{eqnarray}
Here $\ve$ is small positive parameter, $F$ is operator of
the perturbation, the  values of the parameter $\s=\pm1$
correspond so called focusing or defocusing DS-1
equations.
\par
The equations (\ref{ds1}) at $\ve=0$ describe the
interaction of long and short waves on the liquid surface
if the capillary effects and potential flow are taken into
account \cite{D-S,D-R}. The theorems about the existence
of the solutions for this equations in the different
functional classes are known \cite{Gh-S,Sh}. The inverse
scattering trans\-form method  for the DS equations was
formulated in \cite{Nizhnik}--\cite{F-Sung}. This method
allows  to construct the soliton solutions \cite{B-L-M-P}
and to study the global properties of the solutions, for
instance, the asymptotic behavior at large time
\cite{OK1,OK2}. The asymptotic behavior of the solutions
for nonintegrable DS equations was studied in
\cite{Nakao}.
\par
The perturbed equations (\ref{ds1}) are nonintegrable by
inverse scattering trans\-form method. Explicit form of
the perturbation operator can be defined by small
irregularity of bottom  or by taking into account the next
corrections in more realistic models for the liquid
surface considered in \cite{D-S, D-R}. As main example in
this work we  use the perturbation appearing because of
small irregularity of the bottom: $F\equiv AQ$. Here $A$
is real constant. The sign ($\pm$) of $A$ corresponds to
decreasing or increasing depth with respect to spatial
variable $\xi$. Here we don't discuss the other
perturbations corresponding to the system of liquid
surface though the method we used allows to study the DS-1
equations with more wide class of the perturbation
operators.
\par
The special solution of DS-1 we perturbed here was
constructed in the work \cite{B-L-M-P}:
\bb
q(\xi,\eta,t)={\rho\l\mu\exp(it(\l^2+\mu^2))\over2\cosh(\mu
\xi)\cosh(\l\eta)(1-\s|\rho|^2{\mu\l\over16}(1+\th(\l\eta))
(1+\th(\mu\xi)))},
\label{Q}
\ee
where  $\l,\,\mu$ are positive constants defined by
boundary conditions as $\eta\to-\infty$ and
$\xi\to-\infty$; $\rho$ is free complex parameter of the
solution.
\par
This solution decreases with respect to spatial variables
exponentially. Besides if $\s=1$ and
${\mu\l\over4}|\rho|^2>1$ this solution has singularities
on some  lines $\eta=\const$ and $\xi=\const$.
\par
The solution (\ref{Q}) was called dromion in work
\cite{F-S}. The inverse scattering method for dromion-like
solution of (\ref{ds1}) was developed in \cite{F-S}.
\par
The constructing of the the asymptotic soliton-like
solutions for the per\-tur\-ba\-ti\-ons for the integrable
equations is popular field. The perturbations of the
(1+1)-dimensional integrable equations are investigated in
more detail. In that case the asymptotic solutions are
usable at long time and the Fourier-like integral
corrections were studied in \cite{K}-\cite{OK3}. The
instability of the so\-lu\-ti\-ons of the
(2+1)-dimensional equations with respect to transversal
perturbations was studied for example in
\cite{KP}-\cite{P-S}. But the perturbation of the
essential two-dimensional solutions with respect to
spatial variables of the (2+1)-dimensional integrable
equations is less developed field. First cause is more
diversity of the solutions and second cause is more bulky
formulas for the solutions. Here we must refer the works
\cite{GK1}-\cite{GK3}. In these works the perturbation of
the soliton solution for the DS-2 equations had been
studied. From these results and from the work about  the
asymptotic behavior of nonsoliton solutions for the DS-2
equations \cite{OK1} it follows that the soliton of the
DS-2 equations is instable with respect to small
perturbation of the initial data.  These results stimulate
the studies of the dromion perturbation.
\par
The possibility of the full investigation of the
linearized DS-1 equations plays the main role to construct
the perturbation theory of the nonlinear DS-1 equations.
The linearizations of the integrable equations have basic
set of solutions usually. It allows to solve the
linearized equations by Fourier method. For
(1+1)-dimensional equations it was shown  by  Kaup in
\cite{K1}. The works \cite{K-M}-\cite{sachs} were devoted
to the same questions. The basic sets of the solutions for
the  linearized  DS-2 and DS-1 equations were obtained in
\cite{OK4}, \cite{OK5}.

\section{Problem and result}
\par
We construct the asymptotic solution of equ\-a\-ti\-ons
(\ref{ds1}) on $\mod(O(\ve^2))$ uni\-form\-ly at large
$t$. The perturbation operator is $F\equiv AQ$ and the
boundary conditions for $G_1$ and $G_2$:
\bb
G_1|_{\xi\to-\infty}=u_1\equiv{\l^2\over2\cosh^2(\l\eta)},\quad
G_2|_{\eta\to-\infty}=u_2\equiv{\mu^2\over2\cosh^2(\mu\xi)}.
\label{bc}
\ee
We find the asymptotic solution in the form:
\begin{eqnarray}
Q(\xi,\eta,t,\ve)=W(\xi,\eta,t,\t)+\ve U(\xi,\eta,t,\t),
\nonumber\\
G_1(\xi,\eta,t,\ve)=g_1(\xi,\eta,t,\t)+\ve
h_1(\xi,\eta,t,\t),
\nonumber\\
G_2(\xi,\eta,t,\ve)=g_2(\xi,\eta,t,\t)+\ve
h_2(\xi,\eta,t,\t),
\label{as1}
\end{eqnarray}
where $\t=\ve t$ is slow time. The leading term of the
asymptotics has the form:
$$
W(\xi,\eta,t,\t)=q(\xi,\eta,t;\rho(\t)),
$$
and $g_1,\,\,g_2$ are:
\begin{eqnarray*}
g_1(\xi,\eta,t,\t)=u_1-{\s\over2}\int_{-\infty}^{\xi}d\xi'\pt_{\eta}
|W(\xi',\eta,t,\t)|^2,
\\
g_2(\xi,\eta,t,\t)=u_2-{\s\over2}\int_{-\infty}^{\eta}d\eta'\pt_{\xi}
|W(\xi,\eta',t,\t)|^2.
\end{eqnarray*}
\par
Denote $\g(\t)=1-\s{\mu\l\over4}|\rho(\t)|^2$,
$\g_0=\g(0)$.

{\bf Theorem 1.} {\it If
$$
\g(\t)=\g_0^{\exp(2A\t)},\quad Arg(\rho(\t))\equiv\const,
$$
where $\g_0>1$ at $\s=-1$ and $0<\g_0<1$ at $\s=1$, then
the asymptotic solution (\ref{as1}) with respect to
$\mod(O(\ve^2))$ is useful uniformly at $t=O(\ve^{-1})$.}
\par
{\bf Remark 1.} At more long time $t\ll
\ve^{-1}\log(\log(\ve^{-1}))$ the formulas (\ref{as1}) are
asymptotic solution of (\ref{ds1}) with  respect to
$\mod(o(1))$ only.
\par
The  asymptotic analysis given here is valid for the
solutions  (\ref{Q}) without the singularities. It means
if  $\s=1$, then ${\mu\l\over4}|\rho(\t)|^2<1$ or the same
$\g(\t)>0$. If the coefficient of the perturbation $A>0$,
then $|\rho|$ increases ($\g\to0$ at $\t\to\infty$) with
respect to slow time. It allows to say that the
singularity may appear in the leading term  of the
asymptotics at $\t\to\infty$. However we can't state this
strongly because our asymptotics is usable only at
$\t\ll\log(\log(\ve^{-1}))$, while the order of $\g(\t)$
is greater than $\ve$. The order of a  remainder term
after substituting the asymptotic solution (\ref{as1})
into the equations (\ref{ds1}) is only $o(1)$ at the very
long time $1\ll \t\ll\log(\log(\ve^{-1}))$.
\par
The obtained result for the problem about the interaction
of the long and short waves on the liquid surface shows
that when the depth decreases the the formal asymptotic
solution can be described by adiabatic perturbation theory
of the dromion for the focusing  DS-1 equations at least
for $|t|\ll\ve^{-1}\log(\log(\ve^{-1}))$.
\par
{\bf Remark 2.} The appearance of the singularities in the
solution of non\-in\-teg\-rab\-le cases of  the
Davey-Stewartson equations is known phenomenon
\cite{PSSW}.
\par
{\bf Remark 3.} D. Pelinovsky notes to author, that the
modulation of the parameter $|\rho|$ may  be obtained out
of the "energetic equality" \cite{DP}:
$$
\pt_t\int\int_{\Real^2}d\xi d\eta|Q|^2=
\ve\int\int_{\Real^2}d\xi d\eta(Q\bar F-\bar Q F).
$$

\section{Solution of linearized equations}

\par
Here we remind the formulas for solution of the linearized
DS-1 equations on the dromion  as a background:
\begin{eqnarray}
i\pt_t
U+(\pt_\xi^2+\pt_\eta^2)U+(G_1+G_2)U+(G_1+G_2)Q=iF\\
\nonumber
\pt_\xi V_1=-{\s\over2}\pt_\eta(Q\bar U+\bar Q U),\quad
\pt_{\eta}V_2={-\s\over2}\pt_\xi(Q\bar U+\bar Q U).
\nonumber
\end{eqnarray}
\par
The results of inverse scattering transform \cite{F-S} and
the set  of the basic functions \cite{OK5} are used for
solving of the linearized equations by Fourier method.
However unlike the work \cite{OK5} here the solution of
the DS-1 equations with nonzero boundary conditions is
considered. It leads to the change of the dependency of
the scattering data with respect to time (see also
\cite{F-S}) and of the formulas which define the
dependency  of Fourier coefficients of the solution for
the linearized DS-1 equations in contrast to obtained in
\cite{OK5}.
\par
In the inverse scattering transform one use the matrix
solution of the Dirac  system to solve the DS-1 equation
\cite{Nizhnik}-\cite{F-S}:
\bb
\Bigg(
\begin{array}{cc}
\pt_\xi & 0\\
0 & \pt_\eta
\end{array}\Bigg)
\psi=-{1\over2}\Bigg(
\begin{array}{cc}
0 & Q\\ \s\bar Q & 0
\end{array}\Bigg)\psi.
\label{de}
\ee
\par
Let $\psi^+$ and $\psi^-$ are the matrix solutions of the
Goursat  problem for the Dirac system with the boundary
conditions (\cite{F-S})
\bb
\begin{array}{cc}
\psi^+_{11}|_{\xi\to-\infty}=\exp(ik\eta), & \psi^+_{12}|_{\xi\to-\infty}=0,\\
\psi^+_{21}|_{\eta\to\infty}=0,  & \psi^+_{22}|_{\eta\to-\infty}=\exp(-ik\xi);\\
\psi^-_{11}|_{\xi\to-\infty}=\exp(ik\eta), & \psi^-_{12}|_{\xi\to\infty}=0,\\
\psi^-_{21}|_{\eta\to-\infty}=0, & \psi^-_{22}|_{\eta\to-\infty}=\exp(-ik\xi).
\end{array}
\label{b-psi}
\ee
\par
Denote by $\psi^+_{(j)},\, j=1,2,$ the column  of the
matrix $\psi^+$, then this column is the solution of two
systems.  There are the system (\ref{de}) and  additional
system of time evolution:
\begin{eqnarray}
\pt_t \psi^+_{(1)}=ik^2\psi^+_{(1)}+
i\Big(\begin{array}{cc} 1 & 0\\0 & -1\end{array}\Big)
(\pt_\xi-\pt_\eta)^2\psi^+_{(1)}
\nonumber
\\
+ i\Big(\begin{array}{cc} 0 & Q\\\s\bar Q  &
0\end{array}\Big) (\pt_\xi-\pt_\eta)\psi^+_{(1)}+
\Big(
\begin{array}{cc} iG_1 & -i\pt_\eta Q\\i\s\pt_\xi \bar Q & -iG_2\end{array}
\Big)
\psi^+_{(1)}.
\label{dtpsi}
\end{eqnarray}
One can obtain the equation like this for  the  other
columns of the matrices  $\psi^{\pm}$. These equations are
differ from (\ref{dtpsi}) by the sign of $ik^2$ in first
term of the right hand side only.
\par
Now we write two bilinear  forms defining analogs of the
direct and inverse Fourier transforms. First bilinear form
is
\bb
(\chi,\mu)_f=\int_{-\infty}^{\infty}\int_{-\infty}^{\infty}d\xi
d\eta (\chi_1\mu_1\,\s\bar f\,+\,\chi_2\mu_2\,f).
\label{f1}
\ee
Here $\chi_i$ and $\mu_i$ are the elements of the columns
$\chi$ and $\mu$.
\par
Denote by $\phi_{(1)}$ and $\phi_{(2)}$ the solutions
conjugated to $\psi^+_{(1)}$ and $\psi^-_{(2)}$ with
respect to the bilinear form  (\ref{f1}).
\par
Using  the formulas for the scattering  data (\cite{F-S})
one can write these data:
\bb
s_1(k,l)={1\over4\pi}(\psi^+_{(1)}(\xi,\eta,k),E_{(1)}(il\xi))_Q,
\label{sd1}
\ee
\bb
s_2(k,l)={1\over4\pi}(\psi^-_{(2)}(\xi,\eta,k),E_{(2)}(il\eta))_Q.
\label{sd2}
\ee
Here $E(z)={\hbox{diag}}(\exp(z),\exp(-z))$.
\par
It is shown (\cite{F-S}) the elements of the matrices
$\psi^{\pm}$ are analytic functions with respect to the
variable $k$ as $\pm Im(k)>0$. Using the scattering data
one can write the nonlocal Riemann-Hilbert problem for the
$\psi^-_{11}$ and $\psi^+_{12}$ on the real axes
(\cite{F-S}):
$$
\begin{array}{cc}
\psi^-_{11}(\xi,\eta,k)=\exp(ik\eta)+\exp(ik\eta)\bigg(\exp(-ik\eta)
\int_{-\infty}^{\infty} dl\,
s_1(k,l)\psi^+_{12}(\xi,\eta,l)\bigg)^-,
\\
\psi^+_{12}(\xi,\eta,k)=\exp(-ik\xi)\bigg(\exp(ik\xi)
\int_{-\infty}^{\infty} dl\,
s_2(k,l)\psi^-_{11}(\xi,\eta,l)\bigg)^+.
\end{array}
$$
Here
$$
\bigg(f(k)\bigg)^{\pm}={1\over2i\pi}\int_{-\infty}^{\infty}
{dk'\,f(k')\over k'-(k\pm i0)}.
$$
\par
The Riemann-Hilbert problem  for $\psi^-_{21}$ and
$\psi^+_{22}$ has the form:
\begin{eqnarray*}
\psi^-_{21}(\xi,\eta,k)=
&
\exp(ik\eta)\bigg(\exp(-ik\eta)
\int_{-\infty}^{\infty} dl\,
s_1(k,l)\psi^+_{22}(\xi,\eta,l)\bigg)^-,
\\
\psi^+_{22}(\xi,\eta,k)=
&
\exp(-ik\xi)+\qquad\qquad\qquad\qquad\qquad\qquad\qquad\qquad
\\
&
\exp(-ik\xi)\bigg(\exp(ik\xi)
\int_{-\infty}^{\infty} dl\,
s_2(k,l)\psi^-_{21}(\xi,\eta,l)\bigg)^+.
\end{eqnarray*}

\par
Introduce second bilinear form
\bb
\langle\chi,\mu\rangle_s=\int_{-\infty}^{\infty}\int_{-\infty}^{\infty}dk dl
(\chi^1(l)\mu^1(k)\,s_2(k,l)\,+\,\chi^2(l)\mu^2(k)\,s_1(k,l)),
\label{f2}
\ee
where $\chi^{j}$ is the element of the row $\chi$.
\par
Denote by $\varphi^{(j)},\, j=1,2,$  the row conjugated to
$\psi^{(j)}=[\psi^-_{j1},\psi^+_{j2}]$ with respect to the
bilinear  form (\ref{f2}). Formulate the result about the
de\-com\-po\-si\-ti\-on obtained in \cite{OK5}.

{\bf Theorem 2.} {\it Let $Q$ be such that
 $\pt^\a Q\in L_1\cap C^1$ for $|\a|\le
3$, if $f(\xi,\eta)$ is $\pt^\a f\in L_1\cap C^1$ for
$|\a|\le4$, then  one can  write $f$ in the form
$$
f={-1\over\pi}\langle\psi^{(1)}(\xi,\eta,l),
\varphi^{(1)}(\xi,\eta,k)\rangle_{\hat f},
$$
where
$$
\hat f={1\over4\pi}(\psi^+_{(1)}(\xi,\eta,k),\phi_{(1)}(\xi,\eta,l))_f.
$$}
\par
Using  the theorem 2 one can solve the Cauchy problem for
the linearized DS-1 equations. Here the dependence of
$\hat f$ with respect to $t$ is differ from the same
obtained in  \cite{OK5} because here we use the DS-1
equations with nontrivial boundary conditions (\ref{bc}).
These changes may be obtained using the  results of
\cite{F-S} and \cite{OK5}.

{\bf Theorem 3.}{\it Let $Q$ be the solution of the DS-1
equations with  the boundary  conditions
$G_1|_{\xi\to-\infty}=u_1$ and
$G_2|_{\eta\to-\infty}=u_2,$ and $Q$ satisfy the
conditions of theorem  2, the solution of the first of the
linearized DS-1 equation is smooth and integrable function
$U$ with respect to $\xi$ and $\eta$, where $\pt^\a U\in
L_1\cap C^1$ and $\pt^\a F\in L_1\cap C^1,$ for $|\a|\le
4$ and $t\in[0,T_0]$. Then
\begin{eqnarray}
\pt_t\hat U=i(k^2+l^2)\hat U+\int_{-\infty}^{\infty} dk'
\hat U(k-k',l,t)\chi(k')+
\nonumber
\\
\int_{-\infty}^{\infty} dl'\hat U(k,l-l',t)\kappa (l')+\hat F,
\label{u1}
\end{eqnarray}}

\par
If the boundary conditions in the problem for the DS-1
equation are zero, then $\chi\equiv\kappa\equiv0$. In this
case the formulas of the theorem 3 allow to solve the
linearized DS-1 equation in the explicit form. In this
work  we consider the solution of the DS-1 equation with
nonzero boundary conditions. It leads to integral terms in
the formulas of  theorem 3. In order to solve the
linearized DS-1 equation we must transform the  formula
(\ref{u1}). In the right hand side of (\ref{u1}) the
integral terms are the convolutions. Go over to the
equations for the Fourier transform of the functions $\hat
U(k,l,t,\t)$ with respect to variables $k$ and $l$. As
result we obtain  the linear Schr\"odinger equation:
\bb
i\pt_t \tilde U+(\pt_\xi^2+\pt_\eta^2)\tilde U+
(u_2(\xi\mu)+u_1(\eta\l))\tilde U=\tilde F.
\label{lsh}
\ee
Here
$$
\tilde U(\xi,\eta,t,\t)={1\over2\pi}\int_{\Real^2}dkdl
\hat U(k,l,t,\t)\exp(-ik\eta-il\xi);
$$
$$
\tilde F(\xi,\eta,t,\t)={1\over2\pi}\int_{\Real^2}dkdl
\hat F(k,l,t,\t)\exp(-ik\eta-il\xi).
$$
The same equation without the right hand side was obtained
in \cite{F-S} for the time evolution of the scattering
data for the nonlinear DS-1 equation.
\par
One can obtain the solution of the Cauchy problem with
zero in the initial conditions for the equations
(\ref{lsh}) by the separation of the variables. In our
case the solution of the equations (\ref{lsh}) obtained
by the Fourier method has the form:
\begin{eqnarray*}
\tilde U(\xi,\eta,t)=
{1\over2\pi}\int_{\Real^2}dmdn\breve U(m,n,t)X(\xi,m)Y(\eta,n)
\exp(-it(m^2+n^2))+\\
{1\over\sqrt{2\pi}}
\int_{\Real}dn\breve U_{\mu}(n,t)Y(\eta,n)X_{\mu}(\xi)\exp(-it(n^2-\mu^2))
+\\
{1\over\sqrt{2\pi}}
\int_{\Real}dn\breve U_{\l}(m,t)X(\xi,m)Y_{\l}(\eta)\exp(-it(m^2-\l^2))
+\\
\breve U_{\mu,\l}X_{\mu}(\xi)Y_{\l}(\eta)\exp(it(\mu^2+\l^2)).
\end{eqnarray*}
Here we use notations:
$$
X(m,\xi)={\mu\th(\mu \xi)+im\over im-\mu}\exp(-im\xi),
\quad X_{\mu}(\xi)={1\over2\ch(\mu\xi)};
$$
$$
Y(n,\eta)={\l\th(\l \eta)+in\over in-\l}\exp(-in\eta),
\quad Y_{\l}(\eta)={1\over2\ch(\l\eta)};
$$
\begin{eqnarray*}
\pt_t\breve U=i(m^2+n^2)\breve U+\breve F(m,n,t),
\quad
\pt_t \breve U_{\mu}=i(n^2-\mu^2)\breve U+\breve F_{\mu}(n,t),\\
\pt_t \breve U_{\l}=i(m^2-\l^2)\breve U+\breve F_{\l}(m,t),
\quad
\pt_t \breve U_{\mu\l}=-i(\l^2-\mu^2)\breve U+\breve F_{\mu\l}(t),\\
\breve U|_{t=0}=\breve U_{\mu}|_{t=0}=\breve U_{\l}|_{t=0}=
\breve U_{\mu\l}|_{t=0}=0;
\end{eqnarray*}
\begin{eqnarray*}
\breve F(m,n,t)=\int_{\Real^2}d\xi d\eta \tilde F(\xi,\eta,t)
\bar X(m,\xi)\bar Y(n,\eta),\\
\breve F_{\mu}(n,t)=\int_{\Real^2}d\xi d\eta \tilde F(\xi,\eta,t)
\bar X_{\mu}(\xi)\bar Y(n,\eta),\\
\breve F_{\l}(m,t,\t)=\int_{\Real^2}d\xi d\eta \tilde F(\xi,\eta,t)
\bar X(m,\xi)\bar Y_{\l}(\eta),\\
\breve F_{\mu\l}(t,\t)=\int_{\Real^2}d\xi d\eta \tilde F(\xi,\eta,t)
\bar X_{\mu}(\xi)\bar Y_{\l}(\eta).
\end{eqnarray*}

\section{The equation for first correction}
\par
In this  part the equation for the slow modulation of the
parameter $\rho(\t)$ is obtained. This equation is
necessary and sufficient condition for the uniform
boundedness of the  first correction of the expansion
(\ref{as1}).
\par
Substitute  the formula (\ref{as1}) into the equations
(\ref{ds1}). Equate the coefficients with the same power
of $\ve$. The equations at $\ve^0$ are realized because
$W,g_1,g_2$ are the  asymptotic  solution of nonperturbed
DS-1 equations. For the first correction we obtain the
linearized DS-1 equations:
$$
i\pt_t U+(\pt_\xi^2+\pt_\eta^2)U+(g_1+g_2)U+(V_1+V_2)W= iH
$$
\bb
\pt_\xi V_1=-{\s\over2}\pt_\eta(W\bar U+\bar W U),\quad
\pt_{\eta}V_2={-\s\over2}\pt_\xi(W\bar U+\bar W U),
\label{lds1}
\ee
where
$$
H=AW-\pt_\t W.
$$
\par
Before to use the formulas from the previously section we
reduce the form of the  right hand side in first of the
equations (\ref{lds1}). In the leading  term  the
parameter $\rho$ depends on the $\t$ only. The other
parameters depend only on the boundary conditions and do
not change under perturbation. For convenience we
represent $\rho(\t)=r(\t)\exp(i\a(\t))$. The derivation of
$W$ with respect to slow variable $\t$ can be written as:
$$
\pt_\t W=\pt_r W  r'+\pt_\a W \a'.
$$
Here the derivatives $r'$ and $\a'$ are unknown. They will
be obtain below.
\par
Compute the function $\hat H$. From the theorem 2 and
formulas of the functions $\psi_{+}$ and $\phi$ (see
Appendix) we obtain:
$$
\hat H(k,l,t,\t)=\exp(-it(\l^2+\mu^2))\bigg(P(k,l;\rho)-
R(k,l;\rho)\pt_\t\rho\bigg),
$$
where
$$
P(k,l;\rho)=\exp(it(\l^2+\mu^2))\widehat{AW},\quad R(
k,l;\rho)=\exp(it(\l^2+\mu^2))\widehat{\pt_\t W}.
$$
In these  formulas we write the dependence on time in
explicit form. It allows to remove the secular terms in
the asymptotic solution (\ref{as1}). The differential
equations for $\breve{U}$ have the forms:
\begin{eqnarray*}
\pt_t\breve U=i(m^2+n^2)\breve U+
\exp(-it(\l^2+\mu^2))(\breve P(m,n;\rho)-\breve R(m,n;\rho)),
\\
\pt_t \breve U_{\mu}=i(n^2-\mu^2)\breve U+
\exp(-it(\l^2+\mu^2))(\breve P_\mu(n;\rho)-\breve R_\mu(n;\rho)),\\
\pt_t \breve U_{\l}=i(m^2-\l^2)\breve U+
\exp(-it(\l^2+\mu^2))(\breve P_{\l}(m;\rho)-\breve R_\l(m;\rho)),
\\
\pt_t \breve U_{\mu\l}=-i(\l^2-\mu^2)\breve U+
\exp(-it(\l^2+\mu^2))(\breve P_{\mu\l}(\rho)-\breve R_{\mu\l}(\rho)).
\end{eqnarray*}
One can see that the secular terms may appear because of
the last term  in the equation for $\breve U_{\mu\l}$. The
requirement of equivalent to zero of this term leads us to
the equation for $\rho(\t)$:
\bb
\breve R_{\mu\l}-\breve P_{\mu\l}=0,\quad
\rho|_{\t=0}=\rho_0.
\label{sec}
\ee
\par
In the result  $\breve U_{\mu\l}\equiv 0$. The other
equations for $\breve U$ are easy to integrate. The
solutions  of this equations are bounded  on the all of
arguments for all time.
\par
We must return to the original of the images $\breve U$ to
state about boundedness of the solution $U, V_1, V_2$ for
the equations (\ref{lds1}). One can  see the direct (from
$U$ into $\breve U$) and inverse (from $\breve U$ into
$U$) integral transforms as the Fourier transform from the
smooth  and exponentially decreasing functions with
respect to the corresponding variables. The Fourier
transform moves such functions into analytic functions
near the real axis. The inverse transform moves these
analytic functions into the exponential decreasing
functions.  So the solution of (\ref{lds1}) is boundedness
and decreasing exponentially  with respect to the spatial
variables.

\section{Modulation equation for $\rho(\t)$}
\par
Here the equation (\ref{sec}) for the parameter $\rho(\t)$
is reduced to the more  con\-vi\-ent form. Write the
derivative of the leading term with respect to the slow
variable $\t$.
$$
i\pt_\t W=-\a'W- iW{r'\over  r}+{2iW\over 1-\s
r^2{\mu\l\over16}(1+\th(\mu\xi))(1+\th(\l\eta))}\,{r'\over
r}.
$$
Denote $\G={\mu\l\over4}r^2$ and compute the images
$\breve{(\cdot)}_{\mu\l}$ of every term.
$$
\breve{(W)}_{\mu\l}={\s\bar\rho\exp(-it(\l^2+\mu^2))\over 8}
(\s\G-1)\log|1-\s\G|;
$$
$$
\breve{(iW)}_{\mu\l}=-i\s\G{\rho\exp(-it(\l^2+\mu^2))\over 8}.
$$
Denote the image of the last term by $\breve h_{\mu\l}$.
Its image has the form:
$$
\breve h_{\mu\l}={r'\over r}{\s\bar\rho\exp(-it(\l^2+\mu^2))\over8}
(1-\s\G)\bigg({1\over 1-\s\G}-1-\log|1-\s\G|\bigg).
$$
The image $\breve{(\cdot)}$ of $AW$ has the similar form.
\par
Substitute these formulas into ({\ref{sec}) and separate
real and imaginary parts of this equation, then
\begin{eqnarray*}
\a'=0,\qquad\qquad\qquad\qquad\qquad\qquad\qquad\qquad\\
{r'\over r}(\s\G-1)\log|1-\s\G|- {r'\over
r}\bigg(\s\G-(1-\s\G)\log|1-\s\G|\bigg)+
\\
A(\s\G-1)\log|1-\s\G|=0.
\end{eqnarray*}
Use  the notation for  $\G$, then the second equation has
the form:
$$
{d\G\over d\t}=-2\s A (1-\s\G)\log|1-\s\G|.
$$
This equation defines the evolution  of the absolute value
of the complex parameter $\rho$. The argument of this
parameter do not change under the per\-tur\-ba\-ti\-on
$F=AQ$.
\par
Solve the equation for $\G$. Denote $\g=1-\s\G$, rewrite
the equation for $\g$, then we obtain:
$$
\g'=2A\g\log(\g).
$$
The solution for this equation  has the form:
$$
\g(\t)=\exp(C\exp(2A\t)),
$$
where $\g|_{\t=0}=\exp(C)$, then we can write the $\g(\t)$
in  the form:
$$
\g(\t)=\g_0^{\exp(2A\t)}.
$$
The theorem 3 is proved.

\vskip3mm
\centerline{\Large Acknowledgements}}
\par
I thank D.Pelinovsky for the discussions of the results
and for the helpful remarks.
\section{Appendix}
\subsection{Explicit formulas}

\par
Here  we remain the explicit forms of the solution for the
Dirac equation with the dromion-like potential. These
forms were obtained in \cite{F-S}. In  our computations we
use first column of the matrix $\psi^+$ only.
$$
\left(\begin{array}{c}
\psi^+_{11}\\
\psi^+_{21}\end{array}\right)
=\left(\begin{array}{c}\exp(ik\eta)\\0\end{array}\right)+
\frac{\int^{\infty}_\eta {dp\l\exp(ikp))\over2\cosh(\l p)}}{
(1-\s|\rho|^2{\l\mu\over16}(1+\tanh(\mu\xi))(1+\tanh(\l\eta)))}
$$
\bb
\times
\left(\begin{array}{c}
{-\s|\rho|^2\l\mu(1+\tanh(\mu\xi))\over8\cosh(\l\eta)}\\
{\s\bar\rho\mu\exp(-it(\l^2+\mu^2))\over2\cosh(\mu\xi)}
\end{array}\right).
\label{psi+1}
\ee

\subsection{Solutions  of  conjugated equations}

\par
Here we  write the problems for the functions conjugated
to $\psi^{\pm}$ with respect to the bilinear forms.
\par
The functions $\phi$ are the solutions of the boundary
problem con\-ju\-ga\-ted to the solutions of the problem
(\ref{de}), (\ref{bc}) with respect to the bilinear form
(\ref{f1}). First column of the matrix $\phi$ is the
solution of the integral equation:
\begin{eqnarray*}
\phi_{11}(\xi,\eta,l,t)=\exp(il\xi)+
{1\over2}\int_{-\infty}^\eta d\eta'
Q(\xi,\eta',t)\phi_{21}(\xi,\eta',l,t),\\
\phi_{21}(\xi,\eta,l,t)=
{-1\over2}\int_{\xi}^{\infty} d\xi' \bar
Q(\xi',\eta,t)\phi_{11}(\xi',\eta,l,t).
\end{eqnarray*}
\par
The explicit formulas for first column of the matrix
$\phi$ used in  section 4 for the dromion potential has
the form:
$$
\left(\begin{array}{c}
\phi_{11}\\
\phi_{21}\end{array}\right)
=\left(\begin{array}{c}\exp(il\xi)\\0\end{array}\right)+
{\int^{\infty}_\xi {dp\mu\exp(ikp))\over2\cosh(\mu
p)}\over
(1-\s|\rho|^2{\l\mu\over16}(1+\tanh(\mu\xi))(1+\tanh(\l\eta)))}
\times
$$
\bb
\times
\left(\begin{array}{c}
{-\s|\rho|^2\l\mu(1+\tanh(\l\eta))\over8\cosh(\mu\xi)}\\
{-\s\bar\rho\l\exp(-it(\l^2+\mu^2))\over2\cosh(\l\eta)}
\end{array}\right).
\label{phi1}
\ee

\par
Second bilinear  form (\ref{f2}) allows to  write the
integral equations conjugated to integral equations which
were obtained from the nonlocal Riemann-Hilbert equation
for the functions $\psi^{\pm}$ in \cite{F-S}. Using this
equations for the functions $\varphi(\xi,\eta,l)$ one can
show that  the functions $\varphi$ are the solutions of
the boundary equations for the Dirac system or the
integral equations:
\begin{eqnarray*}
\varphi_{11}(\xi,\eta,l)=\exp(il\xi)+
{1\over2}\int_{-\infty}^\eta d\eta'Q(\xi,\eta',t)\varphi_{21}(\xi,\eta',l),\\
\varphi_{21}(\xi,\eta,l)=
{1\over2}\int_{-\infty}^\xi d\xi'\bar Q(\xi',\eta,t)\varphi_{11}(\xi',\eta,l).
\end{eqnarray*}
\par
The functions $\varphi$ have the similar form as the
functions $\phi$.

\end{document}